\documentclass[prd,twocolumn,a4paper,amsmath,amssymb]{revtex4} %preprint %twocolumn
\usepackage{graphicx}

\begin{document}
\title{CLeFAPS: Fast Flexible Alignment of Protein Structures \\
Based on Conformational Letters}
\author{Sheng~Wang$^*$}
\affiliation{Institute of Theoretical Physics, Chinese Academy of
Sciences, Beijing 100190, China}

\begin{abstract}
CLeFAPS, a fast and flexible pairwise structural alignment algorithm
based on a {\it rigid-body} framework, namely CLePAPS, is proposed.
Instead of allowing twists (or bends), the {\it flexible} in CLeFAPS
means: (a) flexibilization of the algorithm's parameters through
self-adapting with the input structures' size, (b) flexibilization
of adding the aligned fragment pairs (AFPs) into an one-to-multi
correspondence set instead of checking their position conflict, (c)
flexible fragment may be found through an elongation procedure
rooted in a vector-based score instead of a distance-based score. We
perform a comparison between CLeFAPS and other popular algorithms
including rigid-body and flexible on a closely-related protein
benchmark (HOMSTRAD) and a distantly-related protein benchmark
(SABmark) while the latter is also for the discrimination test, the
result shows that CLeFAPS is competitive with or even outperforms
other algorithms while the running time is only 1/150 to 1/50 of
them.

{\bf Contact*:} wangsheng@itp.ac.cn

\end{abstract}
\maketitle
%=====================================================================

\section{Introduction}
The comparison of protein structures has been an extremely important
problem in computational biology for a long time \cite{main1ref},
and has been employed in almost all branches of contemporary
structural biology \cite{main2ref1}, where two categories of
application can be achieved from the result of pairwise alignment of
protein structures \cite{MATT}.

The first category is derived from an exact alignment of
residue-residue correspondences in order to identify the homologous
core, which may be called {\it alignment} problem. It can be applied
to make the functional prediction \cite{function1}, to construct
benchmark datasets on which sequence alignment algorithms can be
tested \cite{sequence1}, to discover sequence-structure-motif that
enables protein structure prediction \cite{prediction1}. Finding the
optimal correspondences that are structurally similar between the
two input proteins has been proved to be NP-hard \cite{nphard1}.
However, a practical solution can be obtained by first finding the
local similar fragment pairs (SFPs) between two proteins with a
certain similarity metric and then piling up those SFPs with a
certain consistency metric \cite{DALI,CE}. For example, CLePAPS
\cite{CLePAPS} searches for SFPs with conformational letters
\cite{cle1,cle2} and afterwards applies a ProSup-like \cite{ProSup}
procedure. These algorithms treat protein structures as {\it
rigid-bodies}, while the followings treat them as {\it flexible}
\cite{FATCAT,MATT}. Proteins are flexible molecules that undergo
significant structural changes as part of their normal function
\cite{flexible3}. However, for those current algorithms which
introduce {\it flexibility}, the principal method is allowing twists
(bents), regardless of whether these bents are meaningful or
meaningless \cite{MATT}. Moreover, it has been demonstrated that for
a certain case (drawing ROC curve), the rigid version of FATCAT
outperforms the flexible one \cite{topsfatcat}. Finally, it has been
shown that the runtime of these algorithms is some bit slow
\cite{MATT,FlexProt}.

The second category is derived from a scoring function for the
assessment of the pairwise protein structures' similarity based on
an exact or fuzzy alignment, which may be called {\it assessment}
problem. It can be applied to give a Yes/No answer to distinguish
between 'alignable' and 'non-alignable' proteins \cite{MUSTANG}, to
classify the known protein structures into hierarchical system
\cite{FSSP,SCOP,CATH}, to search the query protein structure against
a target database \cite{search3}. The classical geometric way is the
length of alignment (LALI) plus the root mean squared deviation
(RMSD). Clearly, this is a bi-criteria optimization problem where
the goal is to minimize the RMSD while maximizing the number of
residues \cite{kabsch2}. However, since the RMSD weights the
distances between all residue pairs equally, a small number of local
structural deviations could result in a high RMSD, even when the
global topologies of the compared structures are similar. More
assessment functions have been suggested
\cite{SIscore,LGscore,MaxSub} while these functions have only solved
the first problem by providing a single assessment score while the
other problem is the dependence of the score magnitudes on the
evaluated proteins' size \cite{TMscore}.

Just as the user of a sequence alignment program can control the
'gappiness' by adjusting gap penalties, changing parameters can make
the structural alignment method handle different purposes,
\cite{ProSup} gave a suggestion for parameter settings to deal with
distantly-related proteins, other algorithms optimize a best
parameter set on a training group for general purposes
\cite{CE,TMalign}. However, if the alignment task (for example, the
database search) contains different types of proteins, such as
closely-related, distantly-related, small size and large size, it
will incur inaccuracy or ineffectiveness when assigning fixed
parameters.

We proposed a new approach called CLeFAPS that introduces {\it
flexibility} based on a {\it rigid-body} framework, namely CLePAPS.
The 'F' in CLeFAPS means, ({\bf a: Self-adaptive strategy})
flexiblization of the algorithm's main parameters through the
incorporation of d$_0$ factor from TM-score \cite{TMscore} to
associate four main parameters with the size of the input proteins;
moreover, combined with {\it seed-explosion} strategy (similar as
BLAST \cite{blast}) for SFP generating, we 'self-adapted' all six
main parameters instead of fixing them to handle different types of
proteins; ({\bf b: Fuzzy-add strategy}) flexiblization in the
pile-up of the alignment through enlargement of one-to-one
correspondence set to one-to-multi which collects all AFPs while
neglecting {\it position conflict} (shown in Fig.
\ref{fuzzyaddfigure}); then applying dynamic programming which uses
TM-score as the objective function to get an optimal alignment path.
(The similar procedure is applied in TM-align through constructing
the TM-score rotation matrix \cite{TMalign}. However, such matrix is
O(n$^2$) space complexity and the following dynamic programming is
again O(n$^2$) time complexity, while CLeFAPS is both O(n) space and
time complexity); ({\bf c: Vect-Elong strategy}) flexible fragment
may be found through the elongation procedure based on the
Vect-score (see Eq. (\ref{vectscore})) to collect {\it local
flexible} fragments (shown in Fig. \ref{vectelongfigure}) after
we've identified two proteins' alignment core where all
residue-residue pairs are within the final distance cutoff. In
addition, the incorporation of TM-score is to solve the second
problem talked above since TM-score is normalized in a way that the
score magnitude relative to random structures is not dependent on
the protein's size \cite{TMscore}.

As a result, for those proteins which are distantly related, the
rigid-body based CLeFAPS is competitive with those algorithms that
allow twists (bents) while the running time is only one percent of
them (see Table \ref{runtimeanalysis}). Moreover, the incorporation
of TM-score has been demonstrated effective by comparing the result
on the discrimination test with LALI+RMSD, while the former got a
nearly 10\% higher true negative rate than the latter (see Table
\ref{RMSDvsTMscore}). Finally we compared CLeFAPS with other three
typical algorithms, namely CLePAPS, CE and MATT, based on their
performances on HOMSTRAD (SCOP family level) \cite{HOMSTRAD} and
SABmark (SCOP superfamily level) \cite{SABmark} while the latter is
also for the discrimination test described in \cite{MATT}. CLeFAPS
is open-source for academic users at [http://....].
%=====================================================================

\section{Method}
\subsection{Notation} \label{notation}
Let {\tt mol1} and {\tt mol2} be two input proteins and {\tt moln1}
and {\tt moln2} be their length, respectively. We simultaneously
transfer each structure to its conformational letter according to
\cite{cle1}, and use {\tt cle1} and {\tt cle2} to indicate.

The output of the pairwise alignment involves an one-to-one
residue-residue correspondence set (we'll call it {\tt ali1} and
{\tt ali2}), an one-to-one AFP correspondence set (may be called
{\tt COR}), a rigid-body transformation (comprising a rotation
matrix R and a translation vector T, we'll call them {\tt ROTMAT}),
a geometric assessment (i.e., LALI+RMSD) and a similarity score
(i.e., TM-score) (shown in Supplementary Fig. 4). Particularly,
one-to-one residue-residue correspondence set means that, given one
position in {\tt mol1}, say {\tt ii}, there at most be one
corresponding position, say {\tt jj} in {\tt mol2}, and they have
the structural similarity correspondence, then we record it as, {\tt
ali1[ii]=jj} and {\tt ali2[jj]=ii}. Given {\tt ali1} (or {\tt
ali2}), we can transfer it to {\tt COR} by extracting every ungapped
contiguous residue-residue pair (we'll call it {\it point-pair} and
use $<${\tt ii,jj}$>$ to indicate it) and vice versa.

Some algorithms, such as CE, use {\it AFP} to describe all local
similar fragment pairs between {\tt mol1} and {\tt mol2} in every
case, including those in the final alignment path and those only
having local similarity. In our algorithm, we divide the original
{\it AFP} into SFP and AFP, where the former is the original meaning
while the latter is a subset of SFP that each AFP should satisfy the
consistency metric, namely cRMS distance cutoff in CLeFAPS. In
details, given {\tt ii} in {\tt cle1}, {\tt jj} in {\tt cle2} and a
range length, we can calculate the CLESUM score \cite{cle1} of the
ungapped fragment pair by the following equation:
\begin{equation}\label{clesum}
score = \sum_{k = 0}^{k<\,len} CLESUM[\,cle1[ii+k]]\,[\,cle2[jj+k]]
\end{equation}

Then we may define a SFP only when its CLESUM score is above a given
threshold. We use SFP({\tt ii,jj;len}) to indicate where {\tt ii,
jj} is the starting position in {\tt cle1} and {\tt cle2} and len is
its range length. Moreover, under a certain {\tt ROTMAT}, a SFP may
become a Full\_{}AFP if every point-pair in the SFP is within a
given distance cutoff, or may become a Part\_{}AFP if there exists a
maximal subset where every point-pair is within the given cutoff and
the number of the subset is at least one. Both Full\_{}AFP and
Part\_{}AFP can be generally called AFP, we may also use AFP({\tt
ii,jj;len}) to indicate. Finally, we'll use pivot\_{}SFP to indicate
the SFP that we use to determine the initial {\tt ROTMAT}.

\subsection{Innovative strategy} \label{innostrategy}
\subsubsection{Self-adaptive strategy} \label{selfadaptive}
The equation of TM-score \cite{TMscore} is as follows:
\begin{equation}\label{TMscore}
TM\mbox{-}score =
\frac{1}{L_N}\sum_{k=1}^{LALI}\frac{1}{1+(\frac{d_k}{d_0})^2}
\end{equation}

where L$_N$ is the smaller length of the input structures, d$_k$ is
the distance between the k-th point-pair of aligned residues, LALI
is the length of the aligned residues and d$_0$ is the factor
associated with the protein size, where:
\begin{equation}\label{d0factor}
d_0 = 1.24 \sqrt[3]{L_N -15} - 1.8
\end{equation}

\begin{itemize}
\item[1).]{\bf Association of d$_0$ with the distance cutoff}\\
\end{itemize}

First we set:
\begin{eqnarray}\label{assocd0[1]}
FIN\_{}CUT = d_0, \nonumber\\
5.0 \le FIN\_{}CUT \le 15.0
\end{eqnarray}
because FIN\_{}CUT is our distance cutoff for evaluating overall
alignment, setting FIN\_{}CUT equals to d$_0$ and using such cutoff
to calculate TM-score means the extraction of those point-pairs
which contribute more than 0.5 to TM-score from the final aligned
correspondence set, and eliminating the remaining. Since we know
when the alignments between two proteins get a TM-score more than
0.5, can we say they belong to the same fold \cite{TMalign}.
Actually, this procedure is similar to MaxSub-score \cite{MaxSub},
only with the difference that MaxSub uses a fixed distance cutoff
d$_0$ by users while ours uses a flexible one by the input
structure's size.

Then we set:
\begin{eqnarray}\label{assocd0[2]}
INI\_{}CUT = 2*d_0, \nonumber\\
5.0 \le INI\_{}CUT \le 15.0
\end{eqnarray}
the INI\_{}CUT is used to construct initial alignment similar as
\cite{CLePAPS}. At the beginning, CLeFAPS only uses a single SFP
(i.e., pivot\_{}SFP) to determine the initial {\tt ROTMAT}, so there
may exist some AFPs that are in the final alignment while under
initial {\tt ROTMAT} their point-pairs may still have a large
distance. In order to add these AFPs, we have to use a larger
distance cutoff at the beginning and the twofold scaling is well for
different purposes (see Result \S\ref{different}). The similar
strategy that using a larger INI\_{}CUT than FIN\_{}CUT is also
applied by \cite{ProSup}.

We set the lower limit to 5.0{\AA} for the reason that, if we set
the lower limit below 5.0{\AA}, when dealing with small and
distantly related proteins, the algorithms will miss some
point-pairs which should be in the final alignment (see Result
\S\ref{smallprotein}). While we set the lower limit at 5.0{\AA} to
deal with closely related proteins, the result is still correct.

We set the upper limit to 15.0{\AA} because, while d$_0$=15.0{\AA},
the corresponding length is about 2500 residues (see Eq.
(\ref{d0factor})), this value is nearly the size limit of a single
domain. Moreover, the distance between two adjacent C$_\alpha$ atom
is about 3.8{\AA}, so 15.0{\AA} is about four C$_\alpha$'s length
that when a point-pair's distance is beyond this value may we
basically say they do not have obvious structural correspondence.

\begin{itemize}
\item[2).]{\bf Association of d$_0$ with the average CLESUM score's
threshold}\\
\end{itemize}

Compared to the above part, the association of d$_0$ with CLESUM
score's threshold is arbitrary, we use the following equation,
\begin{eqnarray}\label{assocd0[3]}
THRES\_L = d_0  - 5.0, \nonumber\\
0 \le THRES\_L \le 10
\end{eqnarray}
\begin{equation}\label{assocd0[4]}
THRES\_{}H=15+THRES\_{}L
\end{equation}

Particularly, we set the range of THRES\_{}L from 0 to 10 is
reasonable, since the purpose to create SFP\_{}L is for {\it
sensitivity} which means the list will cover as many SFPs as
possible so that it won't exclude any one that should be in the
final alignment \cite{CLePAPS}. If one SFP gets a similarity score
more than 0, may we say they have the local similarity compared to
the background. For large proteins, however, if we still fix the
threshold at 0, there'll be too many SFPs that make the algorithm
ineffective. When setting the boundary at 10, we may get reasonable
result while reducing 30\% of the running time compared to fixing at
0 (see Result \S\ref{largeprotein} for details).

The reason why we set the range of THRES\_{}H from 15 to 25 is as
follows, since the purpose to create SFP\_{}H is for {\it
specificity} which means the list will contain SFPs with high enough
similarity for constructing an initial {\tt ROTMAT}, while excluding
many purely local coincident SFPs \cite{CLePAPS}. Then, the average
CLESUM score of 15 is high enough to collect highly similar SFPs.
For the same reason as THRES\_{}L, setting the boundary at 25 will
gain effectiveness while retaining accuracy for large proteins.

\subsubsection{Fuzzy-add strategy} \label{fuzzyadd}

%-----------
% Figure.2
%-----------
%[Fig. \ref{fuzzyaddfigure} almost here]\\

\begin{itemize}
\item[1).]{\bf Fuzzy-add}\\
\end{itemize}

\begin{figure}[htb]
\centering
\includegraphics[width=0.72\columnwidth]{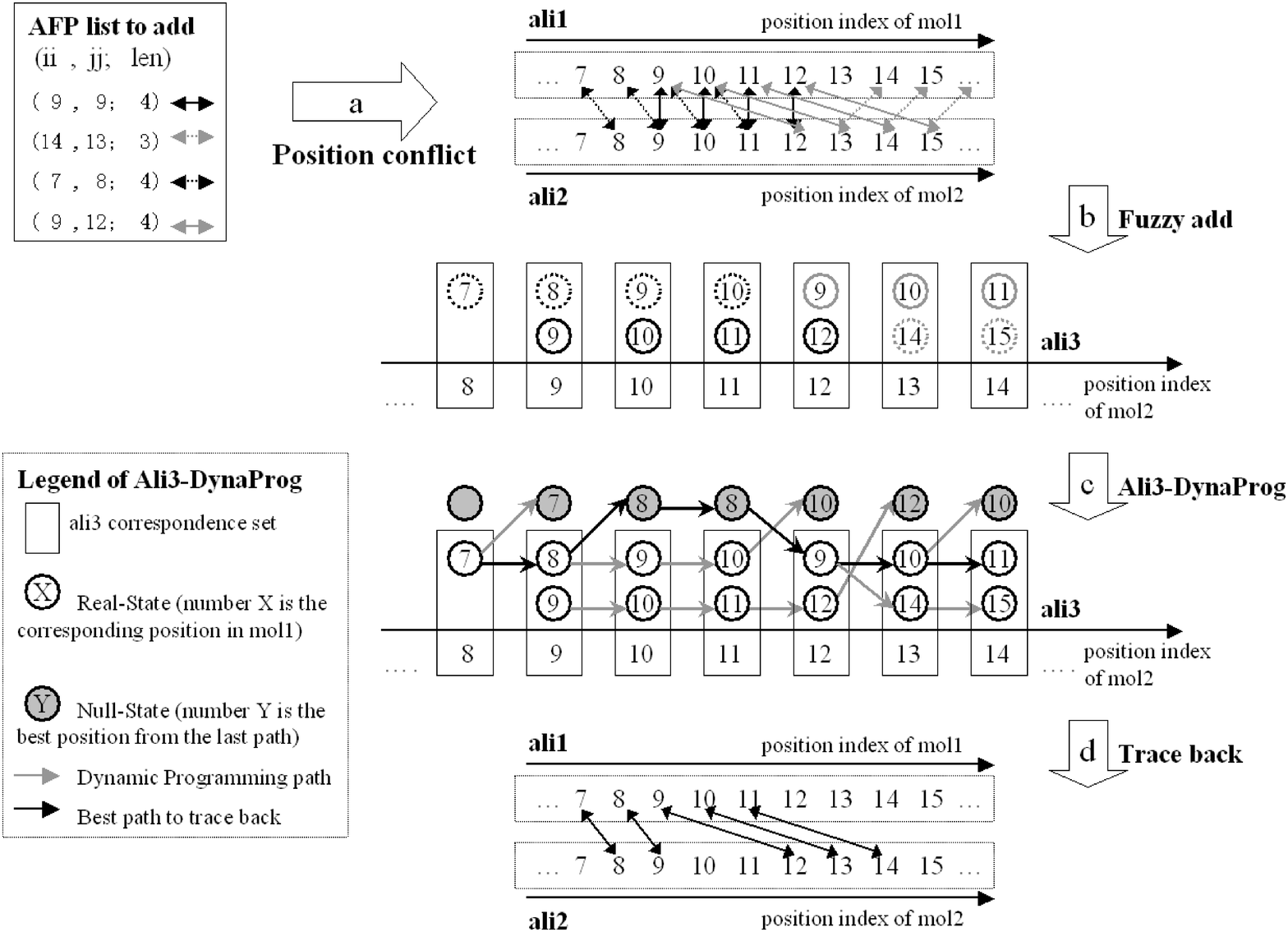}
\caption{An idealized example of the {\it fuzzy-add} strategy. (a)
Suppose the current AFPs to add are in AFP list, if we directly add
them to one-to-one correspondence set, there will occur {\it
position conflict}. (b) If we extend {\tt ali2} to {\tt ali3} with
the second dimension limit (ali3\_{}TOT) set to 2, then we may
directly add AFPs without considering their position conflict. (c)
We may use dynamic programming to find an optimal path which
maximizes a given score. (d) The optimal path can be traced back,
which is just our one-to-one correspondence set.}
\label{fuzzyaddfigure}
\end{figure}

Suppose the AFP list to add is all within the distance cutoff under
a certain {\tt ROTMAT} (actually it contains Full\_{}AFP and
Part\_{}AFP). Then at {\tt ali1} and {\tt ali2}, there will occur
{\it position conflict} (shown in Fig. \ref{fuzzyaddfigure}(a)) that
one position in {\tt mol2} may have more than one corresponding
positions in {\tt mol1}.

A reasonable solution is to extend our one-to-one correspondence
set, say {\tt ali2}, to the one-to-multi set, say {\tt ali3}. The
first dimension in {\tt ali3} is the same as in {\tt ali2} which is
just the position index of {\tt mol2}, while at a given index, the
second dimension is the corresponding position in {\tt mol1} (shown
in Fig. \ref{fuzzyaddfigure}(b)). When adding AFPs, we just need to
put all of them into {\tt ali3}, without having to consider their
{\it position conflict}. This is the definition of fuzzy-add.

In addition, the default value of the maximal number (ali3\_{}TOT)
of the second dimension in {\tt ali3} is 6, that is to say, given
one position in {\tt mol2}, we only consider at most 6 corresponding
positions in {\tt mol1}. When there appears more than 6 positions,
we'll pop-out the position with maximal distance. During AFP adding,
there is only a very small proportion of positions in {\tt mol2}
that will have more than 6 corresponding points. That is because the
maximal distance cutoff in our algorithm is 15.0{\AA} (average is
about 8.0{\AA}), which is about 3 to 4 (average is about 2 to 3)
C$_\alpha$-C$_\alpha$'s distance.

\newpage
\onecolumngrid

\twocolumngrid

\begin{itemize}
\item[2).]{\bf Ali3-DynaProg}\\
\end{itemize}

The purpose of alignment is to get an one-to-one correspondence set
between two proteins, and a natural method that converts
one-to-multi to one-to-one is dynamic programming \cite{DynaProg}
(see Fig. \ref{fuzzyaddfigure}(c)). In details, we design three
temporary data structures, called sco3, pos3 and pre3, to record the
best score through the dynamic programming path, the best position
associated with 'Null-State' (see below) and the traceback pointer,
respectively. Their first dimension is just the same as {\tt ali3},
however the second dimension is one more than {\tt ali3}, the extra
state is called 'Null-State' which deals with gaps (shown in Fig.
\ref{fuzzyaddfigure}(legend)).

\begin{itemize}
\item[] {\tt Ali3-DynaProg:}

\begin{scriptsize}
\begin{ttfamily}
Recursion: for(i=0; i$<$moln2-1; i++)\\
-------------------------------------------------\\
01] for(x=0; x$\le$N[i+1]; x++)\{\\
02] \ \ \ \ if(x==0)\{ // Null-State\\
03] \ \ \ \ \ \ \ \ sco3[i+1][x] = MAX(k=0; k$\le$N[i]; k++)\{\\
04] \ \ \ \ \ \ \ \ \ \ \ \ sco3[i][k] \};\\
05] \ \ \ \ \ \ \ \ pos3[i+1][x] = pos3[i][k\_{}max];\\
06] \ \ \ \ \ \ \ \ pre3[i+1][x] = k\_{}max; \}\\
07] \ \ \ \ else\{ \ \ \ \ // Real-State\\
08] \ \ \ \ \ \ \ \ sco3[i+1][x] = MAX(k=0; k$\le$N[i]; k++)\{\\
09] \ \ \ \ \ \ \ \ \ \ \ \ sco3[i][k]+ \\
10] \ \ \ \ \ \ \ \ \ \ \ \ GAP\_{}FUNCTION(i+1, x; i, k)+ \\
11] \ \ \ \ \ \ \ \ \ \ \ \ SCORE\_{}FUNCTION(i+1, x) \};\\
12] \ \ \ \ \ \ \ \ pos3[i+1][x] = ali3[i+1][x];\\
13] \ \ \ \ \ \ \ \ pre3[i+1][x] = k\_{}max; \}\}\\

\end{ttfamily}
\end{scriptsize}
\end{itemize}
{\tt N[k]} is the total corresponding points of {\tt ali3[k]}, less
than ali3\_{}TOT. {\tt k\_{}max} is the k that maximizes the MAX
function. This is the main dynamic programming function, where,

\begin{itemize}
\item[] {}
\begin{scriptsize}
\begin{ttfamily}
01] GAP\_{}FUNCTION(i+1, x; i, k)\{\\
02] \ \ \ \ cur\_{}pos=ali3[i+1][x]; \ // current position at mol1\\
03] \ \ \ \ bak\_{}pos=pos3[i][k]; \ \ \ // last position at mol1\\
04] \ \ \ \ if(cur\_{}pos$>$bak\_{}pos+1)\{ \ \ \ \ \ // sequential gap\\
05] \ \ \ \ \ \ \ \ return FOR\_{}GAP+(cur\_{}pos-bak\_{}pos)*EXTEND; \}\\
06] \ \ \ \ else if(cur\_{}pos==bak\_{}pos+1)\{ // no gap\\
07] \ \ \ \ \ \ \ \ return 0; \}\\
08] \ \ \ \ else\{ \ \ \ \ \ \ \ \ \ \ \ \ \ \ \ \ \ \ \ \ \ \ // non-sequential gap\\
09] \ \ \ \ \ \ \ \ return BAK\_{}GAP; \}\}\\

01] SCORE\_{}FUNCTION(i+1, x)\{\\
02] \ \ \ \ ii=ali3[i+1][x]; \ \ \ \ // position at mol1\\
03] \ \ \ \ jj=i+1; \ \ \ \  \ \ \ \  \ \ \ \ \ // position at mol2\\
04] \ \ \ \ score =\\
05] \ \ \ \ \ \ \ \ weight1*TM-score(ii,jj) +\\
06] \ \ \ \ \ \ \ \ weight2*Vect-score(ii,jj);\\
07] \ \ \ \ return SCALE*score;\}\\

\end{ttfamily}
\end{scriptsize}
\end{itemize}

There is an important result needed to point out, though dynamic
programming is applied, we may still get non-sequential alignment.
This is because the path of Ali3-DynaProg is sequential to {\tt
mol2}, regardless of the corresponding position in {\tt mol1}.
However, we know that such situation will not often happen, so we
set non-sequential gap penalty a relatively more negative value than
sequential one, in order to punish the former.

\begin{itemize}
\item[]

\begin{flalign}\label{tmscore}
\begin{split}
TM\mbox{-}score(ii, jj)
\end{split}&
\end{flalign}

\begin{scriptsize}
\begin{ttfamily}
01] p1 = mol1[ii];\\
02] p2$'$ = mol2[jj];\\
03] p2 = ROTMAT*p2$'$;\\
04] tm\_{}score = 1.0/(1.0+($|$p1-p2$|$/d$_0$)$^2$);\\

\end{ttfamily}
\end{scriptsize}

\begin{flalign}\label{vectscore}
\begin{split}
Vect\mbox{-}score(ii, jj)
\end{split}&
\end{flalign}

\begin{scriptsize}
\begin{ttfamily}
01] v1 = mol1[ii]-mol1[ii-1];\\
02] v2$'$ = mol2[jj]-mol2[jj-1];\\
03] v2 = ROTMAT*v2$'$;\\
04] vect\_{}score = 1.0*dot(v1,v2)/ ($|$v1$|$ * $|$v2$|$);\\

\end{ttfamily}
\end{scriptsize}
\end{itemize}
the range of TM-score({\tt ii,jj}) is from 0.0 to 1.0, while
Vect-score({\tt ii,jj}) is from -1.0 to 1.0. We arbitrarily set the
Ali3-DynaProg's parameters as follows: SCALE = 100, BAK\_{}GAP =
200, FOR\_{}GAP = 50, EXTEND = 5, and it works well.

After we've applied the Ali3-DynaProg, the optimal path that
maximize the score can be traced back, which is automatically
transferred to an one-to-one correspondence set, that is {\tt ali1}
and {\tt ali2} (shown in Fig. \ref{fuzzyaddfigure}(d)). The
computation time of Ali3-DynaProg grows as O(ali3\_{}TOT*{\tt
moln2}).

\subsubsection{Vect-Elong strategy} \label{vectelongstrategy}
\begin{itemize}
\item[1).] {\bf Vect-score}
\end{itemize}

%-----------
% Figure.3
%-----------
%[Fig. \ref{fragdislocation} almost here]\\
\begin{figure}[htb]
\centering
\includegraphics[width=0.8\columnwidth]{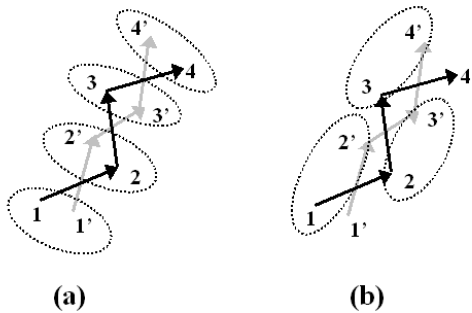}
\caption{An idealized example of {\it fragment dislocation}
misalignment situation, point (1,2,3,4) in dark belongs to {\tt
mol1}, point (1',2',3',4') in gray belongs to {\tt mol2}. (a) The
fragment dislocation misalignment, with point-pairs $<$1,1'$>$,
$<$2,2'$>$, $<$3,3'$>$ and $<$4,4'$>$. (b) The reasonable alignment,
with point-pairs $<$1,2'$>$, $<$2,3'$>$ and $<$3,4'$>$.}
\label{fragdislocation}
\end{figure}

If we measure two protein structures' alignment only rooted in its
point-pair's Euclidean distance, then the situation called {\it
fragment dislocation} misalignment (see Fig. \ref{fragdislocation})
is likely to happen, where the fragment aligned by a certain
algorithm does not stay at its best location, especially in
beta-sheet, with one to four residues deviation.

For example, even the alignment's RMSD may be relatively low in Fig.
\ref{fragdislocation}(a), it's not as reasonable as the alignment
illustrated in Fig. \ref{fragdislocation}(b), where the C$_\beta$
residues are in the same orientation. Such measuring method based on
Euclidean distance may be called 'Dist-score' (e.g., cRMS, TM-score,
etc).

An effective solution to the above problem is to introduce an extra
measuring method, called 'Vect-score' (see Eq. (\ref{vectscore})).
Based on Vect-score, the alignment in Fig. \ref{fragdislocation}(b)
will certainly get a higher score than the alignment in Fig.
\ref{fragdislocation}(a). ProSup also finds a similar example and
applies a different strategy called 'C$_\beta$ filter' to eliminate
such cases.

\begin{itemize}
\item[2).] {\bf Vect-Elong}
\end{itemize}

%-----------
% Figure.4
%-----------
%[Fig. \ref{vectelongfigure} almost here]\\
\begin{figure}[htb]
\centering
\includegraphics[width=0.5\columnwidth]{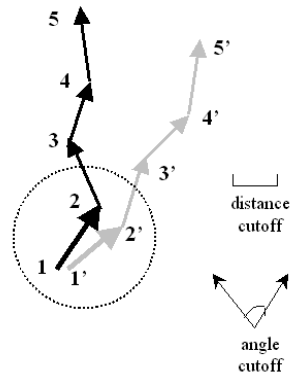}
\caption{An idealized example of {\it Vect-Elong} strategy, point
(1,2,3,4,5) in dark belongs to {\tt mol1}, point (1',2',3',4',5') in
gray belongs to {\tt mol2}. After we have identified two proteins'
alignment core which all point-pairs in the core are within the
distance cutoff as in the figure (we use the dotted circle to
indicate them), we'll of course miss the SFP(3,3';3) that satisfies
the {\it local flexible} condition. If we then apply Vect-Elong with
the angle cutoff in the figure as our parameter, the SFP(3,3';3)
will be added to the correspondence set.} \label{vectelongfigure}
\end{figure}

Another important usage of Vect-score is to deal with the {\it local
flexible} situation (Fig. \ref{vectelongfigure}) defined as follows,
when we have identified two proteins' alignment core which all
point-pairs in the core are within the distance cutoff, there may
exist an AFP (Full\_{}AFP or Part\_{}AFP) outside the core which
meets the following two features: (a) one terminal of the AFP, whose
distance is within the distance cutoff, while the other terminal is
beyond the cutoff; (b) the AFP's corresponding point-pairs are on
basically the same direction.

Vect-Elong is such a procedure to solve {\it local flexible} that
based on Vect-score. We starting from the refined correspondence
set, checking one of these corresponding point-pairs (for example,
$<${\tt ii,jj}$>$) whether or not can be extended to blank portion
(i.e., none of the position in point-pair $<${\tt ii+1,jj+1}$>$ has
corresponding ones). If the point-pair to be tested is blank, and
its Vect-score is within a given threshold, we will add this
point-pair to the correspondence set, and the extension continues.

Fig. \ref{vectelongfigure} for example, if we use an extension
procedure simply based on Dist-score, then for a given distance
cutoff, $<$3,3'$>$, $<$4,4'$>$, $<$5,5'$>$, these three point-pairs
with obvious local similarity will not be added. While we apply
Vect-Elong (using angle cutoff), all these three point-pairs now can
be added.

\subsection{The flowchart and details of CLeFAPS} \label{flowchart}

%-----------
% Figure.5
%-----------

An overview of CLeFAPS is presented in Fig. \ref{overviewfigure}
(see Supplementary Table II for default parameters). Though the
framework is similar as CLePAPS \cite{CLePAPS}, the details of every
step is totally different (see Supplementary for algorithmic
comparison).

\subsubsection{SFP generating} \label{SFPgenerate}
We use {\it seed-explosion} strategy to generate two lists of SFPs.
The seed-explosion strategy is similar as BLAST \cite{blast}, where
we first seek short SFPs at a given length (LEN\_{}L) and a minimal
threshold (THRES\_{}L) (we may call these short SFPs {\it seed}),
then we extend the seed at both terminals. The similar strategy is
also used by MUSTANG and MATT to create their SFPs, while the
difference is that MUSTANG uses cRMS as their similarity metric and
the extension (only at the C-termini) won't stop until the
similarity metric is below the given threshold \cite{MUSTANG}, MATT
also uses cRMS but their SFP's length is from 5 to 9 \cite{MATT}.

We set an extension limit (LEN\_{}H) and a threshold (THRES\_{}L)
for SFP\_{}L. Then we check the extended SFP's average score is more
than THRES\_{}H or not, if it passes the check, we start a second
extension phase whose extension limit is 2*LEN\_{}H and the
threshold is THRES\_{}H to create SFP\_{}H. The extension phase
stops either the current SFP's average score is below the given
threshold, or it's length is beyond the extension limit. After
generating these two lists, we sort them by CLESUM score,
respectively. This step grows O(w1*n$^2$), where n is the longer
protein length, and w1=LEN\_{}H+LEN\_{}L. In real program we use
redundancy shaving procedure that we only keep the SFP with the
highest score among the nearby SFPs \cite{CLePAPS}. (For details of
the pseudo code, see Supplementary, the same as follows.)

We recommend to set the parameters above as follows, LEN\_{}L=6,
LEN\_{}H=9. So the SFP\_{}L is from 6 to 8, and the SFP\_{}H is from
9 to 18. Length 6-8 is necessary for including most SFPs with local
similarity, while length 9-18 will exclude as many SFPs that only
have local coincidence as possible, especially in helix regions
whose average length is about ten \cite{CLePAPS}.

%[Fig. \ref{overviewfigure} almost here]\\

\subsubsection{Select the best pivot\_{}SFP} \label{bestpivot}
We select the best pivot\_{}SFP from TopK of SFP\_{}H according to
its TM-score calculated by fuzzy-adding all AFPs from SFP\_{}H. At
the same time, we get the initial {\tt ROTMAT} according to
\cite{kabsch2}. This step grows O(TopK* SFP\_{}H's size), where the
average space complexity of SFP\_{}H's size is about one third of
SFP\_{}L's and its size is approximately O(1/LEN\_{}H* n$^2$). (See
time complexity analysis in Supplementary.)

We recommend the parameter TopK be 10, that is to say, we'll do at
most 10 recursions to select the best pivot\_{}SFP. This heuristics
is greedy, but it is based on the fact that, if two proteins have
global similarity, the chance that we cannot find one SFP in the
final alignment from the top ten of SFP\_{}H is relatively small.
Actually, our result shows that, at the large database SABmark, the
failure alignment because of this situation (none of top 10 is in
the final alignment) is rare.

\subsubsection{Zoom-in strategy} \label{zoomin}
We apply ZOOM\_{}ITER=3 zoom-in iterations to add AFPs from
SFP\_{}L. First, we use the initial {\tt ROTMAT} from the upper
step; then at k-th iteration we check TopNum of SFP\_{}L for AFPs,
where,
\begin{equation}\label{topnum}
TopNum = \left\{ {\begin{array}{*{20}c}
   {\sum\limits_{i = 1}^k {2^{ - i} } } & {k < ZOOM\_ITER}  \\
   1 & {k = ZOOM\_ITER}  \\
\end{array}} \right.
\end{equation}
meanwhile we gradually lower our distance cutoff by MINUS, where,
\begin{equation}\label{minus}
MINUS= \frac{INI\_{}CUT-FIN\_{}CUT}{ZOOM\_{}ITER}
\end{equation}

For instance, at the first iteration, we check top 1/2 (half) of
SFP\_{}L and the distance cutoff is INI\_{}CUT-MINUS, while at the
final iteration, we check all of SFP\_{}L and the cutoff is
FIN\_{}CUT. At each iteration, we also use fuzzy-add to add AFPs,
then use Ali3-DynaProg to get {\tt COR} which updates {\tt ROTMAT}.
Moreover, we modify the {\it marking} procedure in \cite{CLePAPS}
slightly, if one SFP in SFP\_{}L has none point-pairs within the
distance cutoff, we mark it '-1', then in the later iteration we'll
skip the SFP marked '-1'. This step grows O(ZOOM\_{}ITER*SFP\_{}L's
size) as the worst complexity, however, the introducing of marking
procedure reduces it to O(SFP\_{}L's size).

\subsubsection{Refinement} \label{refinement}
We apply an recursion of maximal REFINE\_{}ITER=10 iterations to
refine our correspondence set under the final distance cutoff
(FIN\_{}CUT), each iteration is constituted by the following three
procedures:

\newpage
\onecolumngrid

\begin{figure}[htb]
\centering
\includegraphics[width=0.8\columnwidth]{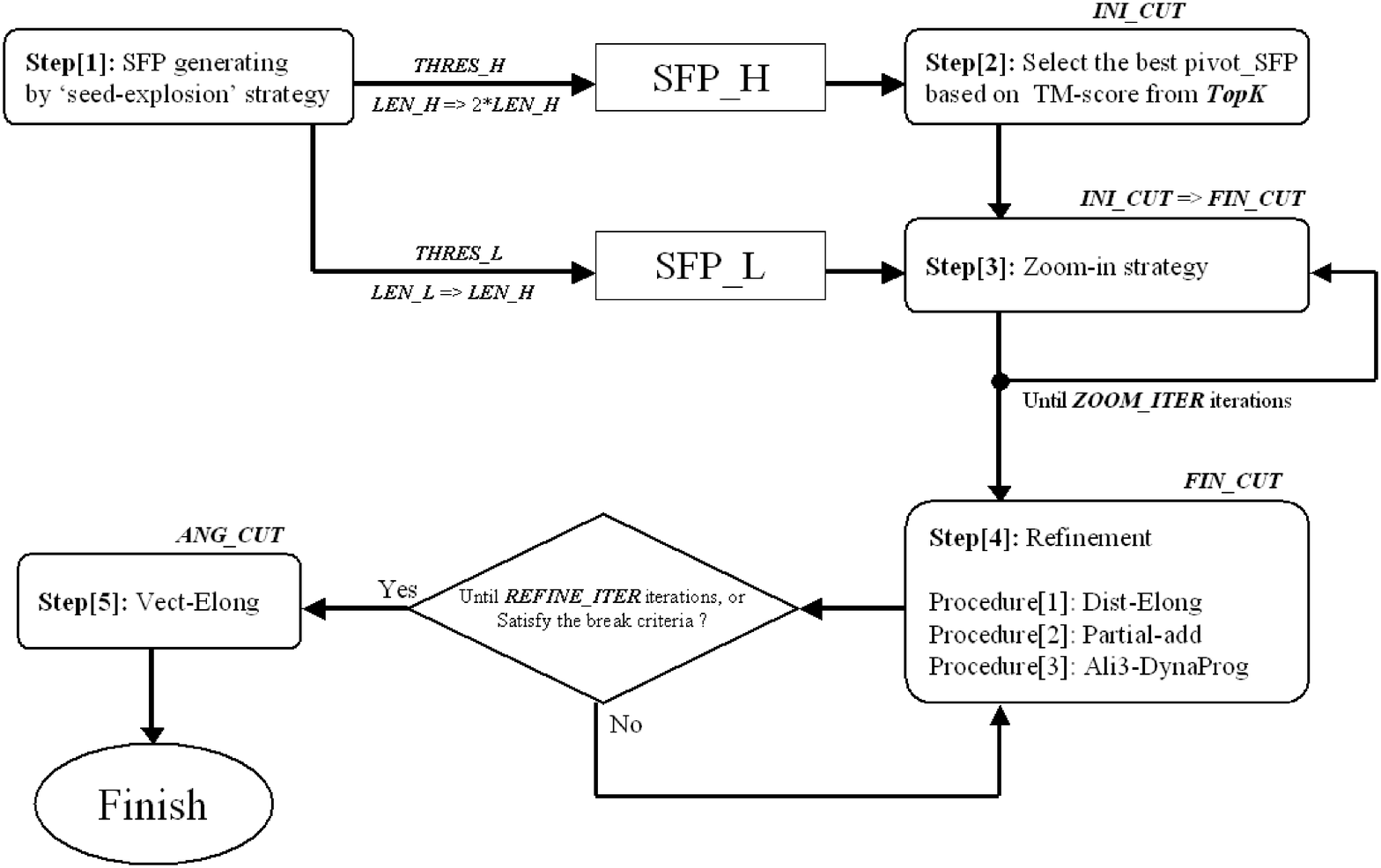}
\caption{An overview of the CLeFAPS algorithm. Words in italic are
the parameters used in related steps.} \label{overviewfigure}
\end{figure}

\twocolumngrid

\begin{itemize}
\item[a).] {\bf Dist-Elong:}
similar as Vect-Elong (see \S\ref{vectelongstrategy}), with the
different that the elongation metric is based on point-pair's
distance instead of Vect-score, and the threshold is FIN\_{}CUT
instead of ANG\_{}CUT.

\item[b).] {\bf Partial-add:}
if one AFP({\tt ii,jj;len}) satisfies the distance cutoff, its
neighbor AFP({\tt ii,jj+k;len}), (where -1*RANGE$\le$k$\le$RANGE)
may also satisfy the cutoff (we call such case {\it partial-move},
there is an excellent illustration in CE's testcase (1col:A with
1cpc:L) \cite{CE} that before and after optimization are obviously
different). So when the {\tt COR} has been identified, we may apply
partial-add to find each AFP's adjacent neighbors, then fuzzy-add
all these AFPs to {\tt ali3}. We set RANGE=4 as default for the
reason that: first, the period of helix is about four C$_\alpha$
residues so it may help to deal with fragment dislocation at helix
region; second, for the other situations except helix, the maximal
distance between four C$_\alpha$'s length is about 15.0{\AA}, which
is near our maximal distance cutoff, beyond which may we basically
say that the point-pair do not have obvious structural
correspondence.

\item[c).] {\bf Ali3-DynaProg:}
different from the above steps (\S\ref{bestpivot} and
\S\ref{zoomin}) which only use TM-score as its SCORE\_{}FUNCTION,
this step uses both TM-score and Vect-score for the purpose to
eliminate fragment dislocation, and setting equal weight works well.
\end{itemize}

At the end of each refinement iteration, we'll apply the following
criteria to check whether to break or not.

\begin{itemize}
\item[] {\tt Break criteria:}

\begin{scriptsize}
\begin{ttfamily}
01] if(Failure\_{}Count $\ge$ FAILURE\_{}CUT)\{ //failure count judge\\
02] \ \ \ \ break; \}\\
03] if(TM\_{}Cur $>$ TM\_{}Max )\{ \ \ \ \ \ \ \ \ \ \ \ //TM-score judge\\
04] \ \ \ \ Failure\_{}Count=0;\\
05] \ \ \ \ TM\_{}Max = TM\_{}Cur;\\
06] \ \ \ \ ROTMAX = ROTCUR; \}\\
07] else if(TM\_{}Cur $<$ 0.95* TM\_{}Max)break;\\
08] else Failure\_{}Count++;\\

\end{ttfamily}
\end{scriptsize}
\end{itemize}
where Failure\_{}Count is the counts of failure that the current
TM-score (TM\_{}Cur) is less than the maximal TM-score (TM\_{}Max),
the default value for FAILURE\_{}CUT is 2, that is to say, if two
continuous recursions cannot make the TM-score better than the
maximal one, we'll break the refinement recursion (this made the
average recursion to about 3-4).

The purpose of the refinement step in our algorithm is similar as in
CE et al. (\cite{CE,ProSup,FATCAT,TMalign,FengSippl}). While the
main difference of ours and theirs is that, CLeFAPS can be run in
O(n) time, however CE et al. use dynamic programming on the distance
matrix calculated using every point-pairs from {\tt mol1} and {\tt
mol2} under current {\tt ROTMAT} so their time complexity is
O(n$^2$).

\subsubsection{Vect-Elong} \label{vectelong}
After refinement, we got the optimized {\tt COR} where every
point-pair is within the final distance cutoff. However, since we
know that there may exist {\it local flexible} situation, it is
recommended to apply Vect-Elong at the final stage with the
parameter ANG\_{}CUT to be 0.6, which will lead to good result.

%=====================================================================

\section{Result}

\subsection{Examples of applying Vect-score and Vect-Elong}
Here, we'll show the following two cases, with comparison of the
four typical algorithms (i.e., CLeFAPS, CLePAPS, CE and MATT, the
same as follows) to show the usage of Vect-score and Vect-Elong.

\begin{figure}[htb]
\centering
\includegraphics[width=0.9\columnwidth]{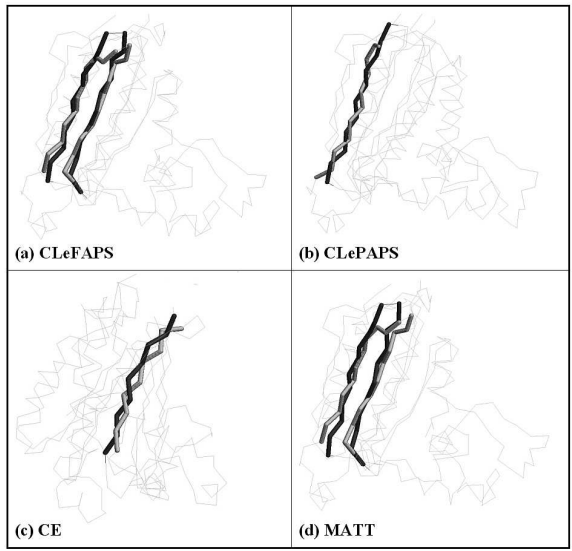}
\caption{Visualization of superposition of dark structure (PDB\_{}ID
1bxd,chain:A,290-450) and gray structure (PDB\_{}ID
1b3q,chain:A,355-540) in HOMSTRAD family {\it Histidine\_{}Kinase}.
Details of the {\it fragment dislocation} misalignment from (b)
CLePAPS and (c) CE; The reasonable alignment from (a) CLeFAPS and
(d) MATT.\\
{\footnotesize Note: residues not placed into the alignment by the
algorithms are shown in thin lines while those in the alignment are
shown in bold lines. The pictures were generated by RasMol
\cite{RasMol}.}}\label{fragdislocationdisplay}
\end{figure}

\subsubsection{Employment of Vect-score to solve {\it fragment dislocation}}

%-----------
% Figure.9
%-----------
%[Fig. \ref{fragdislocationdisplay} almost here]\\

1bxd(chain:A,290-450) and 1b3q(chain:A,355-540) are two protein
domains in the {\it Histidine\_{}Kinase} family of HOMSTRAD. The
fragment dislocation misalignment (in beta-sheet) of CE and CLePAPS
are shown in Fig. \ref{fragdislocationdisplay}(c) and Fig.
\ref{fragdislocationdisplay}(b), respectively. CLeFAPS employs
TM-score plus Vect-score as the SCORE\_{}FUNCTION of Ali3-DynaProg
in the refinement step to eliminate such situation (shown in Fig.
\ref{fragdislocationdisplay}(a)). The result is supported by MATT
(shown in Fig. \ref{fragdislocationdisplay}(d)).

\subsubsection{Employment of Vect-Elong to solve {\it local flexible}}

%-----------
% Figure.10
%-----------
%[Fig. \ref{localflexibledisplay} almost here]\\
\begin{figure}[htb]
\centering
\includegraphics[width=1.0\columnwidth]{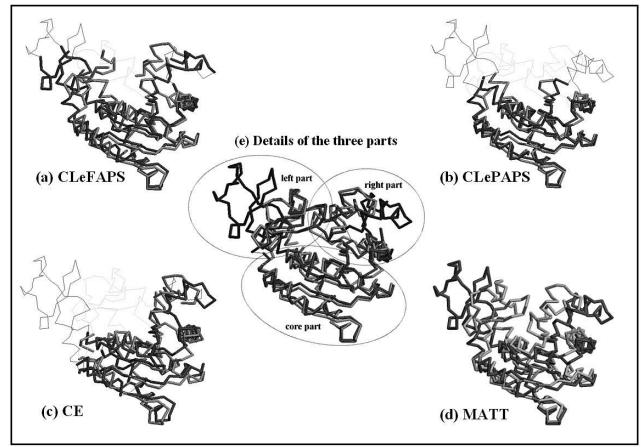}
\caption{Visualization of superposition of dark structure (PDB\_{}ID
1ake) and gray structure (PDB\_{}ID 4ake) aligned by the following
algorithms: (a) CLeFAPS, (b) CLePAPS, (c) CE and (d) MATT. (e)
Details of the three parts: core part, left part and right part.
Cyan structure in left part and right part is the original position
of 4ake, while blue structure in left part and right part is the
individual superposition of 1ake and 4ake based on the
correspondences in left part and right part, respectively.}
\label{localflexibledisplay}
\end{figure}

The adenylate kinase protein (AKE) has a stable inactive
conformation, in addition to an active form, i.e., the open and
closed forms \cite{FlexProt}. They are represented by PDB\_{}ID 4ake
and 1ake, respectively. The protein can be cut into three parts
according to \cite{ake}, which may be called the rigid part (core
part), the LID domain (right part) and the NMP\_{}bind domain (left
part), respectively (shown in Fig. \ref{localflexibledisplay}(e)).
The result alignment of the four algorithms are shown in Fig.
\ref{localflexibledisplay}(a) to \ref{localflexibledisplay}(d).
CLePAPS found the core part, CE found both the core and the right
part, while in right part, CE didn't give an accurate alignment.
CLeFAPS first found the core part similar as CLePAPS, then it
applied Vect-Elong to find the left and the right part, though
incompletely, for the reason that CLeFAPS is based on the rigid-body
framework. MATT did the best job to find all three parts completely,
however it cost the most runtime.

\subsection{Different types of proteins for alignment} \label{different}
We consider the following four different types of proteins, small
size, large size, closely-related and distantly-related. We'll talk
the former two types in this subsection while the latter two types
will be discussed in the following subsection \S\ref{homstrad} and
\S\ref{sabmark}.

\subsubsection{Small proteins} \label{smallprotein}
%-----------
% Figure.11
%-----------
%[Fig. \ref{lowerboundary} almost here]\\
\begin{figure}[htb]
\centering
\includegraphics[width=1.0\columnwidth]{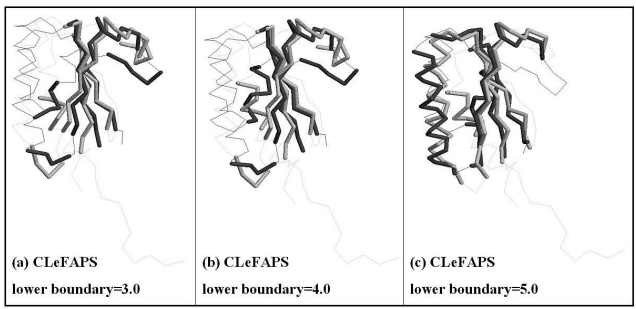}
\caption{Visualization of superposition of d1r5pa\_{} (dark) and
d1t4za\_{} (gray) in SCOP superfamily (c.47.1) aligned by CLeFAPS
with different lower boundary. (a) Lower boundary=3.0, with the
result of TM-score=0.362, LALI=45 and RMSD=2.692. (b) Lower
boundary=4.0 with the result of TM-score=0.376, LALI=48 and
RMSD=2.742. (c) Lower boundary=5.0 (default) with the result of
TM-score=0.442, LALI=63 and RMSD=3.280.} \label{lowerboundary}
\end{figure}

d1r5pa\_{} (90 residues) and d1t4za\_{} (105 residues) belong to the
SCOP {\it Thioredoxin-like} superfamily (c.47.1) (we use the
structures in the ASTRAL(40\%) compendium \cite{ASTRAL}). We've
tried different lower boundary for the association of d$_0$ with the
distance cutoff, from 3.0, 4.0 to 5.0{\AA} (shown in Fig.
\ref{lowerboundary}). The d$_0$ factor in this case is 3.4059{\AA}
(see Eq. (\ref{d0factor})), and the amino acid identity of this pair
is 27.0\%. If we set the lower boundary too small (e.g., 3.0 or
4.0{\AA}), the final distance cutoff (FIN\_{}CUT) will directly be
associated with d$_0$ (see Eq. (\ref{assocd0[1]})), and then we'll
miss some obviously alignable regions when dealing with such small
but distantly related proteins. However, if we set the lower
boundary at a moderate value (i.e., 5.0{\AA}), then when dealing
with proteins whose length is below 180 residues, the final distance
cutoff is constant at 5.0{\AA} (see Eq.
(\ref{d0factor},\ref{assocd0[1]})), and such value is tolerant for
adding AFPs in small (or moderate) size but distantly related
proteins.

\subsubsection{Large proteins} \label{largeprotein}
d1twfb\_{} (1094 residues) and d2a69c1 (1119 residues) belong to the
SCOP {\it beta and beta-prime subunits of DNA dependent RNA-
polymerase} superfamily (e.29.1). We've tried self-adaptive strategy
(association of d$_0$ with the average CLESUM score's threshold) and
constant values. The d$_0$ factor in this case is 10.91{\AA}, using
Eq. (\ref{assocd0[3]},\ref{assocd0[4]}) we get the self-adaptive
threshold (THRES\_{}L=5 and THRES\_{}H=20), then the SFP lists' size
is 5635 of SFP\_{}H and 63644 of SFP\_{}L, respectively. Using
constant values (THRES\_{}L=0 and THRES\_{}H=15) however, the two
SFP lists' size is 10962 of SFP\_{}H and 83737 of SFP\_{}L. As a
result we get the similar alignment with correspondence identity at
94.3\%, while the running time of self-adaptive strategy is 30\%
faster than that of constant values. Moreover, from the comparison
of the four algorithms, CLeFAPS gets the best TM-score 0.720 and the
largest alignment length 851 while the other three get the similar
TM-score (about 0.61) and the similar alignment length (about 700).
This is not surprising, because CLeFAPS employs the final distance
cutoff (FIN\_{}CUT) at 10.91{\AA} so it will collect more alignable
regions than the other algorithms which set their parameters
constant for general purposes instead of such large proteins.

\subsection{CLeFAPS's performance on HOMSTRAD families} \label{homstrad}
HOMSTRAD is a database of protein structural alignments for
homologous families \cite{HOMSTRAD}. Its alignments were generated
using structural alignment programs, then followed by a manual
scrutiny of individual cases. There are totally 1033 families (633
at pairwise level). We'll compare the four algorithms on these 633
families, and the alignment accuracy metric is:

\begin{itemize}
\item[1).] Correct(algorithm)/LOA(length of algorithm)

Calculated by comparing every pairwise alignment in a certain
algorithm against the reference (HOMSTRAD). All correctly aligned
residue pairs in comparison with the reference are considered as
Correct and the total alignment length of the certain algorithm as
LOA. This is the same metric as ACC used in MUSTANG \cite{MUSTANG}.
\end{itemize}

\begin{itemize}
\item[2).] Correct(algorithm)/LOR(length of reference)

All correctly aligned residue pairs in comparison with the reference
are considered as Correct and the length of alignment in reference
is called LOR.
\end{itemize}

The reason why we develop the second C/LOR metric is as follows, for
instance, 1kxr (chain:A,221-352) and 1kfu (chain:L,211-355) are two
protein domains in the {\it Peptidase\_{}C2\_{}D2} family of
HOMSTRAD and reference length is 130. MATT got an alignment of 93
point-pairs with 93 correct, its C/LOA is 1.0 while its C/LOR is
only 0.715. CLeFAPS, however, got an alignment of 123 point-pairs
with 116 correct, its C/LOA is 0.943 while its C/LOR is 0.892.

%===================
% [Table.2]
% Accuracy_Metric
%===================
\begin{table}[htb]
\caption{Alignment accuracy metric on HOMSTRAD from CLeFAPS,
CLePAPS, CE and MATT} \label{accuracyofhomstrad}
\begin{tabular}{c | c  c  c  c}
\hline \hline
Accuracy metric \  &  \ \ CLeFAPS  & \ \ CLePAPS  & \ \   CE    & \ \  MATT \\
\hline
C/LOA$^1$   &   0.929   &   0.916   &   0.911  &  0.948\\
C/LOR$^2$   &   0.898   &   0.847   &   0.881  &  0.831\\
\hline \multicolumn{3}{l}{{\footnotesize $^1$: Correct/Length
of the algorithm.}}\\
\multicolumn{3}{l}{{\footnotesize $^2$: Correct/Length
of the reference.}}\\
\end{tabular}
\end{table}

From the data in Table \ref{accuracyofhomstrad}, MATT scored highest
in C/LOA but lowest in C/LOR. On the contrary, CLeFAPS scored
highest in C/LOR and second highest in C/LOA. This is because MATT
allows local flexibilities (or {\it bent}) everywhere between short
fragments (i.e., AFPs) and then uses dynamic programming to assembly
these {\it bentable} AFPs. However, MATT didn't apply the final
optimization procedure, which is used in CE and CLeFAPS, so the
alignment length of MATT is relatively small while the precision is
relatively high. CLePAPS, analogously, {\it greedy-add} all AFPs and
then skip the final optimization procedure, get a relatively high
C/LOA and low C/LOR as MATT.

\subsection{CLeFAPS's performance on the discrimination problem} \label{sabmark}
The discrimination problem, takes as input a pair of protein
structures, and is supposed to output a yes/no answer (together with
an assessment score) as to whether a good alignment can be found for
these two protein structures or not \cite{MATT}. In our article, we
followed MATT's method and take SABmark \cite{SABmark}'s superfamily
as our test set, which is natural for the discrimination problem
because: (a) it contains 3645 domains sorted into 425 subsets
representing structures at SCOP superfamily level, each SABmark
subset contains at most 25 structures, which can be regarded as {\it
plus} set; (b) it additionally provides a set of decoy structures
for nearly all its 425 sets, each decoy's sequence is similar to its
corresponding set while its structure is within a different SCOP
fold, each decoy set contains at most 25 structures, which can be
regarded as {\it minus} set.

We constructed the following two decoy discrimination test, one is
similar as MATT that for each superfamily we choose a random pair of
structures both from plus set (can not be the same) and a random
pair from plus and minus set, we call such procedure RANDOM test.
The other is that we conduct all-against-all within plus set and
between plus and minus set, we call such procedure All-Against-All
test.

When comparing the four algorithm's ROC curves \cite{ROC}, SABmark
now serves as the gold standard. For varying thresholds based on a
certain assessment function, all pairs below the threshold are
assumed positive, and all above it negative. The pairs that agree
with the standard are called true positives (TP) while those that do
not are false positives (FP) \cite{ROCcurve}.

%========================
% [Table.3]
% LALI+RMSD vs TM-score
%========================
\begin{table}[htb]
\caption{Comparison of LALI+RMSD and TM-score based on MATT's
performance on SABmark} \label{RMSDvsTMscore}
\begin{tabular}{c | c  c }
\hline\hline
True Positive \ & \ \ LALI+RMSD  & \ \ TM-score\\
\hline
 95.04 &   71.16   &   86.0   \\
 94.09 &   75.65   &   87.5   \\
 93.14 &   77.30   &   88.4   \\
 92.20 &   79.20   &   90.3   \\
 91.02 &   82.74   &   91.7   \\
 90.07 &   86.52   &   93.4   \\
\hline \multicolumn{3}{c}{\ \ \ \ \ \ \ \ \ \ \ \ \ \ \ \ \ \ \ \
True Negative}\\
\hline \multicolumn{3}{l}{{\footnotesize Note:}}\\
\multicolumn{3}{l}{{\footnotesize (1) the LALI+RMSD data is from
MATT \cite{MATT}.}}\\
\multicolumn{3}{l}{{\footnotesize (2) the discrimination test is
RANDOM test.}}\\
\multicolumn{3}{l}{{\footnotesize (3)
True\_{}Negative\%+False\_{}Positive\%=100.0\%}}
\end{tabular}
\end{table}

First, we compare the assessment function of TM-score and LALI+RMSD
based on the same algorithm (MATT) and the same decoy discrimination
test (RANDOM test) (see Table \ref{RMSDvsTMscore}) \cite{MATT}, at
each fixed true positive rate, TM-score got a nearly 10\% higher
true negative rate than LALI+RMSD.

%================
% [Table.4]
% AUC_Analysis
%================
\begin{table}[htb]
\caption{AUC values based on TM-score from CLeFAPS, CLePAPS, CE and
MATT} \label{AUCvalue}
\begin{tabular}{c | c  c  c  c}
\hline \hline
Discrimination test \  &  \ \ CLeFAPS  & \ \ CLePAPS  & \ \   CE    & \ \  MATT \\
\hline
RANDOM          &  0.970    &   0.932   & 0.966 & 0.974\\
All-Against-All &  0.952    &   0.912   & 0.956 & 0.964\\
\hline
\end{tabular}
\end{table}

\begin{figure}[htb]
\centering
(a) RANDOM discrimination test\\
\includegraphics[width=1.0\columnwidth]{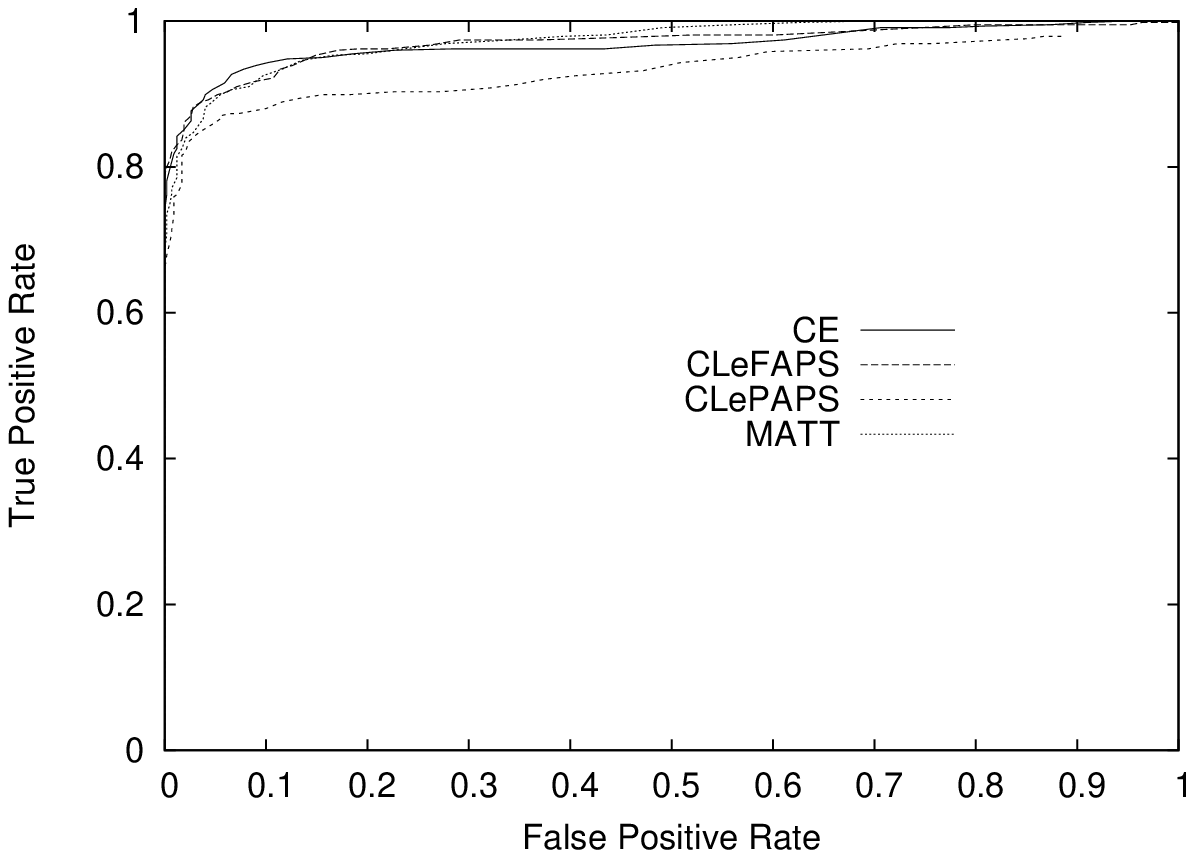}\\
(b) All-Against-All discrimination test\\
\includegraphics[width=1.0\columnwidth]{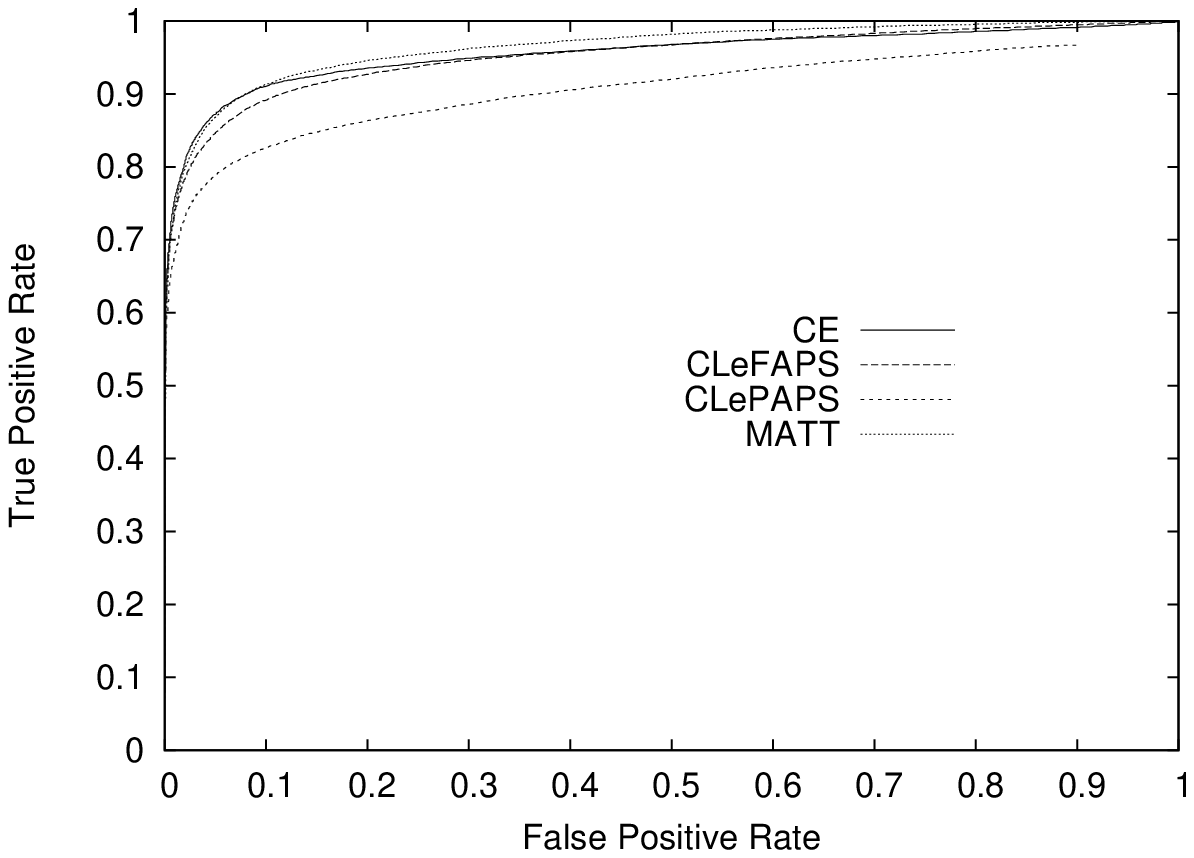}\\
\caption{The ROC curve analysis based on TM-score obtained from
CLeFAPS, CLePAPS, CE and MATT.} \label{roccurvefigure}
\end{figure}

Second, we compare ROC curves and AUC \cite{topsfatcat} over the
four algorithms (shown in Fig. \ref{roccurvefigure} and Table
\ref{AUCvalue}), MATT performs best in both tests and CE follows the
second, while CLeFAPS is comparable with CE and is better than
CLePAPS. A notable result when comparing CLePAPS and CLeFAPS is, in
RANDOM test CLePAPS failed 9/425 in positive test and 49/425
negative test while CLeFAPS only failed 1/425 in the former test; in
All-Against-All test CLePAPS failed 1322/40676 in positive test and
4064/40066 negative test while CLeFAPS failed 28/40676 in former and
75/40066 in latter. This result may be the demonstration that,
CLeFAPS employing the seed-explosion strategy to create SFP\_{}H is
more effective than CLePAPS employing fixed parameters.

%===================
% [Table.5]
% Runtime_Analysis
%===================
\begin{table}[htb]
\caption{Runtime of All-Against-All test from CLeFAPS, CLePAPS, CE
and MATT} \label{runtimeanalysis}
\begin{tabular}{c | c  c  c  c}
\hline \hline
Runtime (sec) \  &  \ \ CLeFAPS  & \ \ CLePAPS  & \ \   CE    & \ \  MATT \\
\hline
Total runtime   &    1259   &   1136    &  61669   & 172812 \\
Average runtime &  0.01526  &  0.01377  & 0.74765  & 2.09510\\
\hline \multicolumn{5}{l}{{\footnotesize Note: All-Against-All test
contains 80742 pairs of proteins.}}
\end{tabular}
\end{table}

Finally, we compare running (see Table \ref{runtimeanalysis}) using
the Windows XP operation system with 2*2.66-GHz Dual-Core Intel CORE
2 Dual processor and 2-GB 667 MHz memory. The result is on
All-Against-All test which contains 80742 pairs of proteins. We find
that, though MATT and CE perform best and second best (comparable
with CLeFAPS) on the discrimination problem, they are the slowest
and the second slowest on running time, while CLeFAPS and CLePAPS
takes only about 1/50 of the running time used by CE and 1/150 of
MATT, and CLeFAPS is only 10\% more than CLePAPS.
%=====================================================================

%------------
% Figure.ROC
%------------
%[Fig. \ref{roccurvefigure} almost here]\\

\section{Discussion and future work}
We proposed the program called CLeFAPS, which considers protein's
flexibility based on a {\it rigid-body} framework, instead of
introducing twists (bends). The result showed that when dealing with
the structural distortion caused by distantly related proteins
through evolution \cite{FATCAT}, CLeFAPS is competitive with those
algorithms that allow twists, and the reasons are as follows,

\begin{itemize}
\item[a).]
Through the incorporation of d$_0$ factor from TM-score to associate
the main parameters of the pairwise alignment, including the
similarity metric of SFP (CLESUM score threshold) and the
consistency metric in pile-up of the alignment (distance cutoff),
with the size of the input proteins ({\it Parameter self-adaptive}).

\item[b).]
Through the enlargement of the one-to-one correspondence set to
one-to-multi during the pile-up procedure, which collects all AFPs
while neglecting their position conflict ({\it Fuzzy-add}). Then
applying dynamic programming, which uses TM-score (or plus
Vect-score) as the objective function, to get an optimal alignment
path ({\it Ali3-DynaProg}).

\item[c).]
Through the elongation based on the Vect-score to collect local
flexible fragments, that the fragment's point-pairs are exceed the
final distance cutoff while they share local structural similarity,
after we've identified two proteins' alignment core ({\it
Vect-Elong}).
\end{itemize}

Furthermore, we employ TM-score as the assessment function to
measure the structural similarity between two proteins, which has
been demonstrated effective by comparing the result on the
discrimination test.

Perhaps the most highlighted feature of CLeFAPS is its fast speed,
where the most important contribution is the TopK(=10) cutoff in the
step called {\it select the best pivot\_{}SFP} (see
\S\ref{bestpivot}), where we'll do at most TopK(=10) recursions. If
all these TopK(=10) SFPs in SFP\_{}H are far away from the final
alignment, the algorithm will certainly end in failure. In the
future work, we'll start a precise exploration on the accuracy of
TopK SFPs in SFP\_{}H through the statistics on some large
databases.

There is another structural distortion caused by conformational
flexibility \cite{FATCAT}, say, domain motion \cite{domainmove}.
However, CLeFAPS is ineffective to deal with such cases because of
its rigid-body framework while it can only deal with {\it local
flexible} fragments. When an entire domain undergoes a significant
conformational change, we may use the {\it Multi-solution} strategy
\cite{CLePAPS,ProSup} to solve it.

CLeFAPS is a sequence-independent structural alignment algorithm,
however if we consider the amino acid, the generalized
conformational letter (reduction of amino acid plus conformational
letter) \cite{cle2} may be employed to encode the input proteins and
the generalized CLESUM \cite{cle2} be applied to generate the SFP
list. It is expected that through this procedure may we get more
accurate result as well as reduce the TopK's failure rate.

\section*{Supplementary Data}
Supplementary Data are available at ....

\section*{Acknowledgments}
We are grateful to professor Wei-mou Zheng, Drs. Ming Li,
Ai-ming Xiong, Kang Li for their helpful discussions, and
colleague Hui Zeng for drawing the ROC curve.

\end{document}